# The Distance to High-Velocity Cloud Complex M

G.L. Verschuur
0000-0002-6160-1040
verschuur@aol.com

J.T. Schmelz
0000-0001-7908-6940
USRA, 425 3rd Street SW, Suite 950, Washington, D.C.
jschmelz@usra.edu

**Abstract**

λ21-cm HI4PI survey data are used to study the anomalous-velocity hydrogen gas associated with high-velocity cloud Complex M. These high-sensitivity, high-resolution, high-dynamic-range data show that many of the individual features, including MI, MIIa, and MIIb, are components of a long, arched filament that extends from about (l, b) = (105º, 53º) to (l, b) = (196º, 55º). Maps at different velocities, results from Gaussian analysis, and observations of associated high-energy emission make a compelling case that the MI cloud and the arched filament are physically interacting. If this is the case, we can use the distance to MI, 150 pc as reported by Schmelz & Verschuur (2022), to set the distance to Complex M. The estimated mass of Complex M is then about 120 solar masses and the energy implied using the observed line-of-sight velocity, -85 km/s, is $8.4 \times 10^{48}$ ergs. Integrating over 4π steradians, the total energy for a spherically symmetrical explosion is estimated to be $1.9 \times 10^{50}$ ergs, well within the energy budget of a typical supernova.

**Introduction**

If the disk of the Galaxy were composed of uniform-density neutral hydrogen (HI) with a thickness of about 200 parsecs, all the high-latitude gas would be local with low velocities due to random motions of order $-20 < v_{LSR} < +20$ km/s. However, observations of anomalous-velocity gas going back to Muller et al. (1963) show that this simple picture of galactic structure is not correct. Subsequently, a division of labor between Leiden and Groningen observatory astronomers led to the traditional separation of high-velocity (HV) gas with $v_{LSR} < -80$ km/s and intermediate-velocity (IV) gas with $-80 < v_{LSR} < -30$ km/s. Such hydrogen clouds and cloud complexes are found all over the sky. They are defined by their observed anomalous velocities, which are inconsistent with the regular rotation of the Milky Way. Please see Verschuur (1975) for an initial review and Wakker et al. (2004) for subsequent developments.

Models of the origin and explanations for the existence of anomalous-velocity features are strongly influenced by distance estimates. Absorption-line studies of both stellar and extragalactic sources traditionally place these clouds at a kiloparsec or more in the galactic halo; see, e.g., the comprehensive surveys compiled by Kuntz & Danly (1996) and Wakker (2001). In principle, a star behind the cloud will show absorption features, but a star in front will not. If the distances to the stars are well-known, the HI cloud should lie between them. This method is illustrated for Complex M by Danly et al. (1993). The upper and lower distance limits are set by the detection and non-detection of IUE spectral features. One star, BD +38 2182, showed corresponding interstellar absorption in strong lines of Si II, C II, and O I with velocities up to -120 km/s. Similar data for HD 93521 showed no absorption lines. Since the two stars are separated by only 27', this method would indicate that Complex M lies at a distance between the two stars, i.e., $1.5 < z < 4.4$ kpc. An optical Ca IIK absorption detection for BD +38 2182 and non-detection for HD 93521 give similar results (Keenan et al. 1995).

These distance results are complicated by pervasive filamentary HI structure (McClure-Griffiths et al. 2009; Peek et al. 2011; Winkel et al. 2016; Martin et al. 2015), which calls into question the assumption of homogeneity over the entire radio beam. Warnings on using a non-detection to constrain the distance were sounded by Ryans et al. (1997). If indeed, the Complex M gas is filamentary, a negative spectroscopy result may indicate that the star's position is between filaments, even for stars that appear close together on the sky. As a result, the traditional absorption-line studies can give us a well-defined upper limit but no useful lower limit. So, in summary, the absorption-line studies can only constrain Complex M to a distance less than several kiloparsecs.

Schmelz & Verschuur (2022) found that the HV cloud MI may be the result of a supernova that took place about 100,000 years ago. Low-velocity hydrogen show a clear cavity centered on the spatial coordinates of MI which coincides with that of a spectroscopic variable yellow giant star, 56 Ursae Majoris. Escorza et al. (2022) have shown that the invisible companion of 56 UMa is a neutron star. The remains of the supernova that created the neutron star may have evacuated the cavity and blasted MI itself outward at the observed high velocity. The distance to the star is 163 ± 7 pc, which allows the distance to MI to be estimated at 150 pc, much closer than the traditional absorption-line studies would indicate. Because the distance is close by astronomical standards, the mass and energy of MI are easily in line with what is expected from a single supernova.

In this paper, we focus our investigations on Complex M, components of which were first observed around (l, b) = (165º, 60º) by Mathewson (1967). We show that several of the individual cloud-like features in Complex M, including MI, form an arch-shaped filament. If MI and this filament are interacting, we can use the results of Schmelz & Verschuur (2022) to estimate the distance to Complex M.

**Analysis**

*Longitude-Latitude Plots*

The overall morphology of the area described in this paper is shown in Fig. 1, which presents maps of galactic longitude (l) versus latitude (b) of the HI brightness at velocities that highlight some of the well-known features of Complex M. The data are from the λ21-cm galactic neutral atomic hydrogen HI4PI survey (Ben Bekhti et al. 2016), which combines the northern hemisphere data from the Effelsberg-Bonn Survey (EBHIS; Winkel et al. 2016) and the southern hemisphere data from the third revision of the Galactic All-Sky Survey (GASS; McClure-Griffiths et al. 2009).

HI4PI has an angular resolution of $\Theta_{FWHM}$ = 16.2 arcmin, a sensitivity of σ = 43 mK, and full spatial sampling, 5 arcmin in both galactic l and b.

The contour intervals in Fig. 1 and subsequent figures are chosen to highlight the morphological details. The naming of various HI structures is traditional, but can be confusing as data quality improves. What appeared to be isolated "clouds" in maps from old surveys might now be seen as parts of filaments as sensitivity, resolution, and dynamic range improve. This may be the case with the well-known HV feature labeled as MI at (l, b) = (165°, 65.°), which looks cloud-like in panel Fig. 1a at a velocity of -120 km/s. Parts of HV Complex C appear in the lower right corner of this panel.

Maps at lower velocities, such as Fig. 1b at -100 km/s, show the connectedness between MI and the filamentary structure that is part of Complex M and includes a loop-the-loop shaped segment, which extends from l = 130° to 160°. In some early papers on HV clouds (e.g., Hulsbosch 1968), both the cloud and the loop-the-loop segment were grouped together. MII is located to the left of MI. Different peaks have been given labels, but those are not always consistent in the literature. We designated two of these, MIIa and MIIb, which appear to extend the complex to greater longitudes. It may be that these features, which appeared as clouds in earlier studies are, rather, local density concentrations or line-of-sight enhancements in a long, twisted filament.

Fig. 1c at -75 km/s clearly shows emission bridging the gap between MI and MIIa & MIIb. Because of the connectedness revealed by the HI4PI data, we are able to follow the arching shape from about (l, b) = (105°, 53.°) to (l, b) = (196°, 55°). We refer to the overall structure as Complex M throughout this paper. In addition, a distinctly different feature, the well-known IV Arch, appears below Complex M. This brings up a challenge related to the modern study of interstellar HI gas. Where clouds had clear boundaries and were isolated on the sky, filaments connect features in both position and velocity space. Defining the boundaries is not necessarily easy. Kalberla & Haud (2006) studied the connection between HV and IV gas in the Complex M. Since this paper presents evidence for a radically different distance, we decided to focus on Complex M where the connections to MI were the strongest, and exclude the IV Arch where the connections to MI were weaker/different/nebulous. It may very well be true that the distance to the IV Arch is much closer than the 0.8 to 1.8 kpc values quoted by Kalberla and Haud (2006), but we have no new evidence (yet) to support this claim. For the purposes of this paper, neither the IV Arch nor Complex C will be considered in the analysis.

Fig. 2 shows detailed l-b maps of Complex M using the EBHIS λ21-cm data, which has a spatial resolution of 10.8 arcmin (Winkel et al. 2016). The panels show how the structure of MI changes with velocity from (a) the familiar two-peaked distribution at -116 km/s to (b) the flattened ring at -108 km/s (Giovanelli et al. 1973). The remaining panels show how this ring slowly morphs from a relatively isolated HV cloud shape to the extended loop-the-loop segment of Complex M. It is not possible to say definitively where the cloud ends and the filament begins, and there is neither a spatial nor a velocity boundary separating the cloud from the filament. On the higher longitude side, however, the connection between MI and MIIa is much more tenuous. We will refer to this as the M&M\ gap. Frames (e) to (h) show that this gap is traversed by a narrow bridge of emission in the velocities from -82 to -70 km/s, implying that whatever created the gap at higher negative velocities did not succeed in removing all the HI at these lower velocities. We will return to the M&M gap in Section 3.

*Longitude-Velocity Plots*

The HI4PI data cube allows us to examine Complex M in different ways. The area maps shown above give us one perspective, but the far-less commonly used longitude-velocity maps give us another view. Fig. 3 shows a series l-v plots at different latitudes that cover the longitude range of interest. The figure shows clear HI ridges at several velocities. The low-velocity gas is generally thought to be relatively close to the Sun, certainly less than 100 pc distant at these latitudes. Normally, this gas is centered at zero velocity over the whole sky, and the offset from zero seen here may be related to the presence of the Local Bubble (Zucker et al. 2022). The second ridge of emission at about -60 km/s forms part of the IV Arch. The third set of peaks around -85 km/s marks the loop-the-loop segment of Complex M.

The panels of Fig. 3 were chosen to highlight the connectedness of MI and the loop-the-loop segment. In Fig. 3a at b = +64°.8, MI appears as an isolated cloud with two main peaks. In Fig. 3b at b = +65°.6, a string of peaks at around -90 km/s begins to appear. They seem to be a series of clouds, but are, rather, enhancements created where the l-v plot at this latitude intersects the twisted structures that define the loop-the-loop segment. This panel also shows the location of MIIb, which marks the continuation of Complex M to higher longitudes as seen in Fig. 1c. Fig. 3c shows the strengthening of the connection between MI and the intensifying peaks of the loop-the-loop segment; the IV Arch is also dominant in this panel. Fig. 3d shows the strongest connection between MI and the lower-longitude gas features. Fig. 3e shows the M&M gap. In addition, HI in the low-velocity ridge at -5 to -10 km/s is missing between l = 162 & 173°. These directions mark the boundaries of the low-velocity cavity created by the MI supernova event discussed by Schmelz & Verschuur (2022). MIIa is brightest in Fig. 3f.

Fig. 3 shows how the connection between MI and the loop-the-loop segment evolves with increasing latitude. The connection starts to form in panel (b) and continues to strengthen throughout the image sequence until they appear to blend (morph) in panel (f). In addition, the data show the M&M gap on the high-longitude side. We see in Fig. 1 that the arched filament appears to be moving coherently across the area which indicates that its axis must be nearly in the plane of the sky. Why then is there an asymmetry, as MI connects strongly with anomalous-velocity gas at lower longitudes but only tentatively to MIIa and MIIb at higher longitudes? The emission from the bridge discussed above is not plainly evident in these l-v plots because of its narrow angular width and relatively low brightness. It appears along the velocity edge of the unrelated emission that peaks around -60 km/s. This highlights our contention that position-position maps and position-velocity maps accentuate different facets of HI structure. We will return to this asymmetry in Section 3.

*Gaussian Analysis*

Additional evidence to support our hypothesis that MI is interacting with Complex M comes from Gaussian decomposition of HI emission profiles. The semi-automated Gaussian fitting method described by Verschuur (2004) uses the Microsoft Excel SOLVER algorithm. SOLVER employs the generalized reduced gradient (GRG2) nonlinear optimization code developed by L. Lasdon, University of Texas at Austin, and A. Waren, Cleveland State University (references to papers published on this method may be found at http:// www.optimalmethods.com). The initialization of SOLVER requires a first guess for the Gaussian parameters, including the approximate center velocity, line width, and amplitude of apparent components until the residuals resemble the noise. SOLVER is then run, adjusting the Gaussians to minimize the residuals and then calculating the reduced chi squared value for the fit. We then examine the solution. If the value of reduced chi squared is between about 0.9 and 1.1, the fit is judged to be satisfactory. If it is too high or too low, we can add or subtract components, respectively, and re-run SOLVER. This process determines the number of individual Gaussians that are required for a final good fit to the data. If a series of closely spaced profiles are to be decomposed, the best fit solution for the first profile is used as the starting point for the next in the series. Only components with brightness temperature of $\geq 0.1$ K, and column densities $\geq 10^{18}$ cm$^{-2}$ were considered in the statistics. We selected two areas of the EBHIS data cube for comparison. The first involved a mapping of MI using profiles every 0.°2 in longitude from 167.°0 to 169.°8 and every 0.°1 in latitude over the range of the HV gas. This sample involved 330 profiles that produced 3182 Gaussians covering all velocities of which 659 pertain to HV HI. A second set of 50 profiles is located along the axis of the relatively straight and narrow section of the Complex M from longitude 140.°0 to 146.°8 and a latitude range that covered the filament. The Gaussian analysis produced 393 components at all velocities of which 68 pertain to HV gas.

Fig. 4 shows examples of the HI profiles for (a) MI and (b) the filament segment. Each profile has multiple peaks representing the complex structure seen is Fig. 3. The right-most spectral feature represents the local gas and the peak at about -60 km/s highlights the IV Arch. Although these portions of the profiles are quite different, it is the left-most peak, with the highest negative velocity, that is of greatest interest to us here. The panels also show the Gaussian fits - one broad component (red) is seen in every direction with one or two weaker narrow components superimposed. The broader components have an average line width of 25.7 ± 3.4 km/s for MI and 24.5 ± 1.5 km/s for the filament segment, thus indicating that the underlying properties of both HV regions are similar.

Gaussian line widths of order 25 km/s are traditionally attributed to a warm neutral or ionized medium with a temperature of about 10,000 K, which might be in pressure balance with both colder and hotter interstellar components. If the gas is this warm, then we might expect to see its spectral signatures at wavelengths other than $\lambda$21-cm. The Wisconsin H$\alpha$ Mapper (WHAM) surveyed the distribution and kinematics of ionized gas in the Galaxy above declination -30° with an angular resolution of 1° (Haffner et al. 2003). The resulting profiles reveal ionized gas in nearly every direction on the sky including many anomalous velocity features at high latitudes.

Fig. 4c compares the HI and H\alpha structure at -80 km/s with a 10 km/s-wide band. The higher HI contour levels are overlaid on the corresponding H\alpha emission in color, and the correlation between the two appears to be significant. A more directed study by Tufte et al. (1998) found clear H$\alpha$ emission in 6 directions toward MI at around -100 km/s, and at two locations toward MIIb at about -80 km/s. Since MI is far from any known source of ionizing radiation, another mechanism is required to heat the gas and produce the observed H$\alpha$ emission. That mechanism may be the interaction between MI and Complex M discussed here.

*High-Energy Emissions*

Herbstmeier et al. (1995) compared X-ray data from the ROSAT 1/4 keV survey with the structure of the Complex M neutral hydrogen. They found an extended, nebulous cloud of X-rays as well as enhanced X-rays toward both MI and MIIb. Blom et al. (1997) report γ-rays detected in the range 0.75-3 MeV with COMPTEL from the area below Complex M arched filament seen in Fig. 1c with a peak at l, b = 165°, 57°. This high-energy emission is often attributed to non-thermal electron bremsstrahlung arising from HV cloud interactions with matter at the halo-disk interface, but since this border is absent in the Local Chimney (see below), a different explanation is required.

The Sun appears to lie in a region of low-density gas known as the Local Bubble, which was first detected using interstellar extinction (Fitzgerald 1968) and Ly-α absorption (Bohlin 1975) measurements toward nearby stars. Pervasive soft X-ray emission (Williamson et al. 1974; Sanders et al. 1977) and O VI absorption towards several solar-neighborhood stars (Jenkins 1978) suggested that the Local Bubble may be filled with a high-temperature, low-density plasma. Observations indicate that the Local Bubble appears to extend to 100-200 parsecs in all directions (Zucker et al. 2022) and is surrounded by an irregular, higher-density gas boundary near the galactic plane. These regions of low-density gas may extend, however, through the disk and into the halo in certain high-latitude directions in a feature known as the Local Chimney (Welsh et al. 1999). The bubble and chimney were mapped in 3D in a series of papers (see Lallement et al. 2003 and references therein) using absorption characteristics of the interstellar NaI D-line doublet at 5890 Å toward over 1000 target stars lying within 350 pc of the Sun as determined by parallax measurements made with the Hipparcos satellite. Fig. 5 shows a cross section of the Local Chimney at longitude 165° from Lallement et al. (2003) highlighting the location of MI and 56 UMa (Schmelz & Verschuur 2022).

These higher-energy emissions suggest that Complex M is doing more than just traveling aimlessly through the cold, remote depths of the galactic halo. In fact, these emissions are all easier to explain if Complex M is much closer to the Sun. Schmelz & Verschuur (2022) showed that the diffuse X-rays reported by Herbstmeier et al. (1995) are located in the lower-left half of the MI cavity where the integrated HI emission is weakest. They also interpreted the enhanced X-rays associated with MI as evidence for a bow shock that forms as the 56 UMa binary system and its supernova cavity travel through interstellar space and interact with ambient gas. Interactions between MI and the arched filament could create the energy needed to ionize hydrogen, generate the Hα emission, and produce the X-rays associated with the denser, fastest-moving features. If the magnetic field is important here, the X-rays could be generated by reconnection or wave heating. The γ-ray emission may not be part of the Complex M arched filament that is the focus of this paper, but it could be yet another manifestation of the process that generated the HV gas and point the way to understanding the energetics of the phenomenon that created Complex M.

**Discussion**

The plots of l-b and l-v, the results from the Gaussian analysis, and the observations of high-energy emission make a compelling case that the MI cloud and the loop-the-loop segment are physically interacting. Although the connection is not beyond a reasonable doubt, it does appear to be in line with the preponderance of evidence. If this is the case, then we can use the distance to the visible component of the single-line spectroscopic binary system, 56 UMa, to bootstrap the distance from MI to Complex M.

*The Distance to Complex M*

The supernova scenario described by Schmelz & Verschuur (2022) to explain the high anomalous velocity of MI was only possible because 56 UMa survived the explosion of its binary companion. They deliberately limited their investigation to explain the origin and energy of MI and resisted the temptation to apply this scenario to other anomalous velocity features where the evidence may be less convincing. Many neutron stars might be hiding in plain sight over much of the sky, creating cavities in the surrounding matter, but without a bright companion, a pulsar signature, or mass transfer, they would be all but invisible to us.

Schmelz & Verschuur (2022) found the radius of MI cavity to about $4°.5$ or 13 pc, far too small to be responsible for the Complex M arched filament seen in Fig. 1. Yet, the previous section presents compelling evidence that the MI cloud and the loop-the-loop segment are interacting. Fig. 1 shows that the arched filament appears to be moving coherently at a velocity of about -85 km/s. It may be longer than it appears, but this length is what we can see easily when we look up the Local Chimney because there is little intervening gas to confuse the structure. We propose the following scenario - the arched filament is created and launched by an unknown mechanism (see Section 3.3). Then, about 100,000 years ago, the invisible companion of 56 UMa exploded, blasting MI away at a velocity of 120 km/s. Recently, at least astronomically speaking, MI caught up with and began interacting with/crashing into/passing through the arched filament. If this scenario is correct, that would place Complex M at the same distance as MI - about 150 pc.

With this distance, we can estimate the mass and energy of Complex M. Based on a composite HI spectrum from $l = 135°$ to $165°$ and $b = 65°$ to $68°$, it is possible to estimate the average column density for the loop-the-loop segment of Complex M to be about $20.8 \times 10^{18}$ cm$^{-2}$. Making the standard assumption that the depth is about equal to the width of the filament ($0°.5$), the volume density is 5.3 cm$^{-3}$, the mass is about 40 solar masses, and the energy is $2.8 \times 10^{48}$ ergs. If the original source for the motion of Complex M were a supernova, this derived mass would be a combination of the matter from the original explosion as well as gas swept up via the snowplow effect (Spitzer 1978) as the blast wave moves through interstellar space. Without knowing the original radial velocity of this hypothetical supernova progenitor, we can only estimate the energy required to accelerate the matter from zero velocity. The full extent of Complex M may be three times greater than just the loop-the-loop segment, implying a mass of 120 solar masses and an energy of $8.4 \times 10^{48}$ ergs. These values require the same assumptions as those described above. Integrating over $4\pi$ steradians, the total energy for a spherically symmetrical explosion is estimated to be $1.9 \times 10^{50}$ ergs, well within the energy budget of a typical supernova ($10^{51}$ ergs).

*The Asymmetry of Complex M*

If MI and the loop-the-loop segment are physically interacting, the supernova that created the MI cavity would have ionized the ambient HI so that the M&M gap in the Complex M ridge would be centered on MI. But the emission expected from Complex M across this region is only missing between the longitude of MI, 165º, and 175º and not at longitudes less than 165º. Why this asymmetry? This phenomenon, which we have referred to as the M&M gap, is clearly visible in Figs. 1-3. Fig. 1c shows a thin filament that bridges the gap, which has either survived the impact of the supernova shell or the shell has not yet reached it.

Why, then, does the supernova shell centered on MI produce this asymmetric pattern of missing gas in the overall morphology of Complex M? Fig. 6 offers a possible explanation. This asymmetry can be accounted for if the axis of the loop-the-loop segment, which is otherwise fairy straight, is inclined along our line-of-sight. Overall, its orientation may reflect the modulating influence of the local interstellar magnetic field. Radio polarization vectors at 408 MHz (Brouw & Spoelstra 1976) and optical polarization data for stars (Matheson & Ford 1970) show that the local interstellar magnetic field is normal to the line-of-sight at $l = 140º$.

Fig. 6 shows schematically and to scale the relationship between the MI cavity carved out by the 56 UMa-related supernova and its interaction with the arched filament. Its overall axis is likely to be curved in depth as well in the plane of the sky. The supernova shell interacts with the loop-the-loop segment at longitudes greater than that of MI at $l = 165º$, to remove the ambient neutral hydrogen gas that would otherwise be found at -85 km/s. Instead, the M&M gap is created. According to the geometry in Fig. 6, the gas at longitudes less than 165º has not yet been affected by the expanding shell. Beyond longitude 175º the HI gas in Complex M is again relatively undisturbed.

*The Origin of Complex M*

Although the connection to MI allows us to bootstrap the distance to Complex M, its origin remains a mystery. There are several leading contenders for the responsible mechanism, including an old, local supernova, which is an obvious candidate. This model, which was rejected by Oort (1966) based on what was known about anomalous-velocity gas at the time, was resurrected by Schmelz & Verschuur (2022) to explain MI. The supernova that created the cavity centered on the 56 UMa binary system is too small have produced the entire arched filament (Fig.1), but another larger, older explosion may be responsible. The energy calculated above make this scenario possible.

Another mechanism that could be responsible for the origin of Complex M is the galactic fountain. Observations from early rocket-borne X-ray detectors revealed wide-spread diffuse high-energy emission (Williamson et al. 1974). These X-rays were thought to originate from hot gas produced by supernovae in the disk that could escape into the halo. Shapiro & Field (1976) suggested a convective-radiative galactic fountain model to cool and condense hot halo gas as it fell back toward the galactic plane. Bregman (1980) then examined the kinematics and found good agreement with the observed velocities of HV features.

A third possible mechanism is the Perseus super-shell. Verschuur (1993) modeled the velocities and spatial distribution of Complex M and other nearby anomalous velocity features as part of a vast super-shell with an elliptical cross section normal to the galactic disk. Its origin appears to be in the Perseus arm toward l = 131±4º with an apparent radius of 1920 pc and an expansion velocity of 281±10 km/s. The Perseus arm is home to dozens of OB associations, making multiple supernovae a possible origin for the subset of anomalous velocity gas features in the second quadrant of the northern galactic hemisphere. According to that model, gas in the near side of the shell has impacted a local structure marked by the radio continuum spur known as Loop III. The derived distances of the anomalous velocity HI driven by the super-shell are dependent on location and velocity, ranging from 100 pc or less for gas in the near face to as much as a kpc.

**Conclusions**

Using λ21-cm galactic neutral atomic hydrogen data from the HI4PI survey (Ben Bekhti et al. 2016), we have examined the anomalous velocity gas features in the second quadrant of the northern galactic hemisphere associated with Complex M. Much of this gas, including the well-known features designated MI, MIIa, and MIIb, are part of a long, arched filament extending from about (l, b) = (105º, 53.º) to (l, b) = (196 º, 55.º).

Plots of l-b and l-v show that the MI cloud and the loop-the-loop segment of Complex M are physically interacting. This view is supported by results from Gaussian analysis and observations of high-energy emission. These interactions could create the energy needed to ionize hydrogen, generate the H\alpha emission, and produce the X-rays associated with the more-dense, fastest-moving features.

Schmelz & Verschuur (2022) showed that MI is located at the edge of a cavity surrounding the single-line spectroscopic binary system, 56 UMa, whose distance is 163 pc. That allowed them to derive the distance of MI as 150 pc. The physical connection to the loop-the-loop segment thus allows us to bootstrap the distance to Complex M, whose distance is then of order 150 pc.

This result conflicts with past estimates that place all HV gas complexes in the galactic halo, but a new understanding of the filamentary nature of interstellar HI gas has helped establish the view that absorption line data can be used only to establish an upper limit to the distance to the HV hydrogen. The new distance determined for Complex M is consistent with these results.

For a distance of 150 pc, the mass of Complex M is about 120 solar masses. If the original source were a supernova, this mass would be a combination of the matter from the explosion as well as gas swept up as the blast wave moved through interstellar space.

Without the radial velocity of this hypothetical supernova progenitor, we can only estimate the energy required to accelerate the matter from zero velocity. This energy is $8.4 \times 10^{48}$. Integrating over 4π steradians, the total energy for a spherically symmetrical explosion is estimated to be $1.9 \times 10^{50}$ ergs, well within the energy budget of a typical supernova ($10^{51}$ ergs).

The asymmetric connection between MI and the arched filament, with the M&M gap on the left and the loop-the-loop segment on the right, is explained by the orientation of Complex M, which curves not only in the plane of the sky, but also along the line of sight (Fig. 6). The arched filament appears to be following the local magnetic field direction.

The distance to Complex M, which is nearby by astronomical standards, allows for the possibility that the anomalous-velocity gas is the result of an old, local supernova. But without a bright binary companion, a pulsar signature, or mass transfer, the resulting neutron star would be all but invisible to us.

Without the neutron star to nail down the supernova model, other explanations like the galactic-fountain (Shapiro & Field 1976; Bregman 1980) and super-shell model (Verschuur 1993) are still very much in play for the origin of Complex M and the larger-scale distribution of anomalous-velocity HI.

## Acknowledgments

We are grateful to T. Dame for providing us with his MacFits software that seamlessly allows us to unravel data cubes, to J. Kerp for providing the HI4PI data, and to B. Winkel for sending us the EBHIS data.

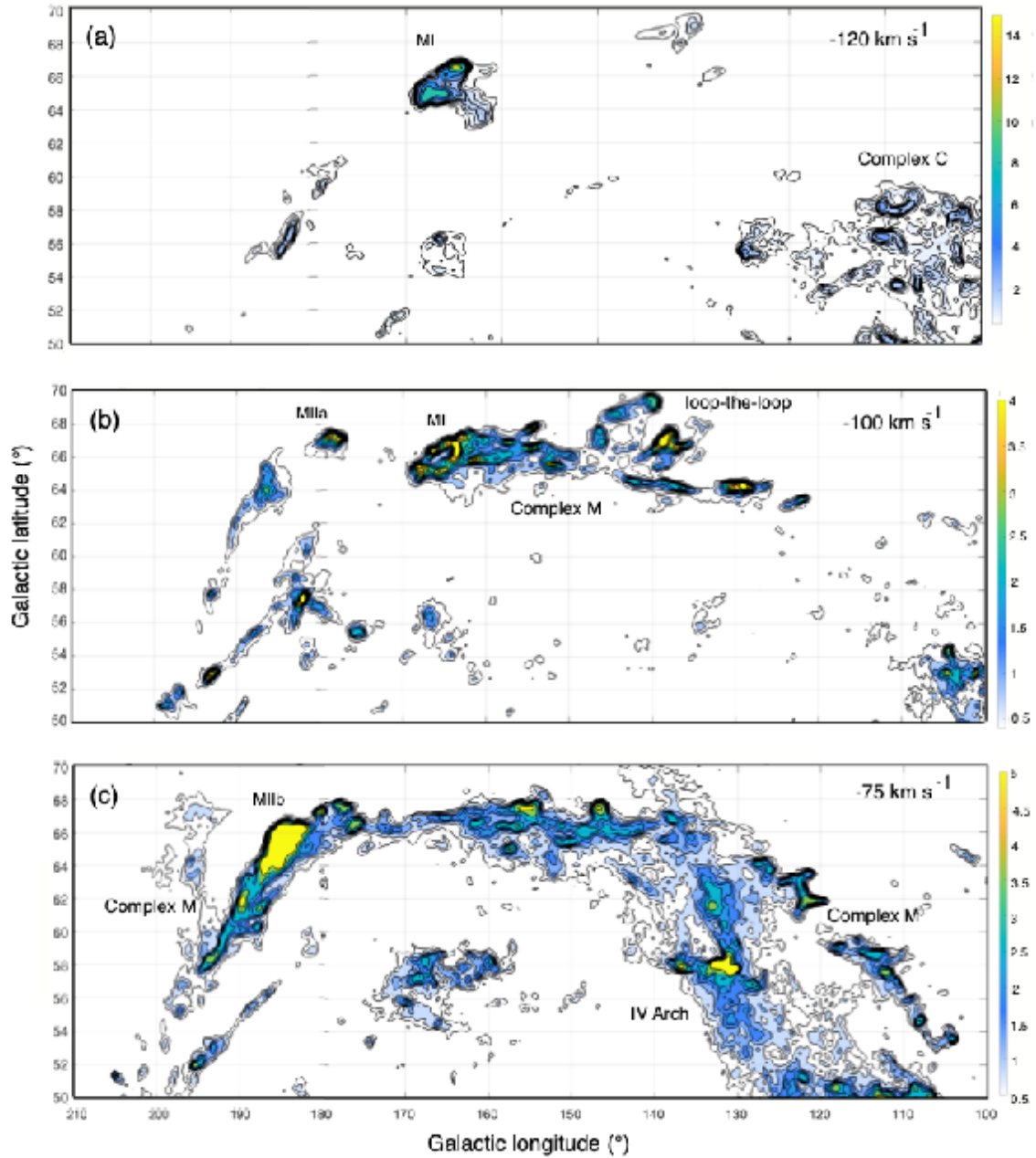

Fig 1. Area maps of the HI emission from Complex M produced using HI4PI data at the velocities indicated in a bandwidth of 2 km/s. (a) At -120 km/s, MI dominates and parts of HV Complex C appear in the lower-right corner of the frame. (b) At -100 km/s, MI is seen to be part of filamentary structure that includes the loop-the-loop segment centered at l = 140º. Cloud-like MIIa appears to the left of MI and forms part of an extension to higher longitudes seen in (c) at -75 km/s. Here, emission bridges the gap between MI and MIIa to form an overall, long, arching, twisted filament that can be followed from about (l, b) = (105º, 53º) to (l,b) = (196º, 55). Also visible here is the IV Arch which appears to be a distinctly different feature. Contour levels in this and subsequent figures are chosen to highlight morphological details. Legends are in degrees K.

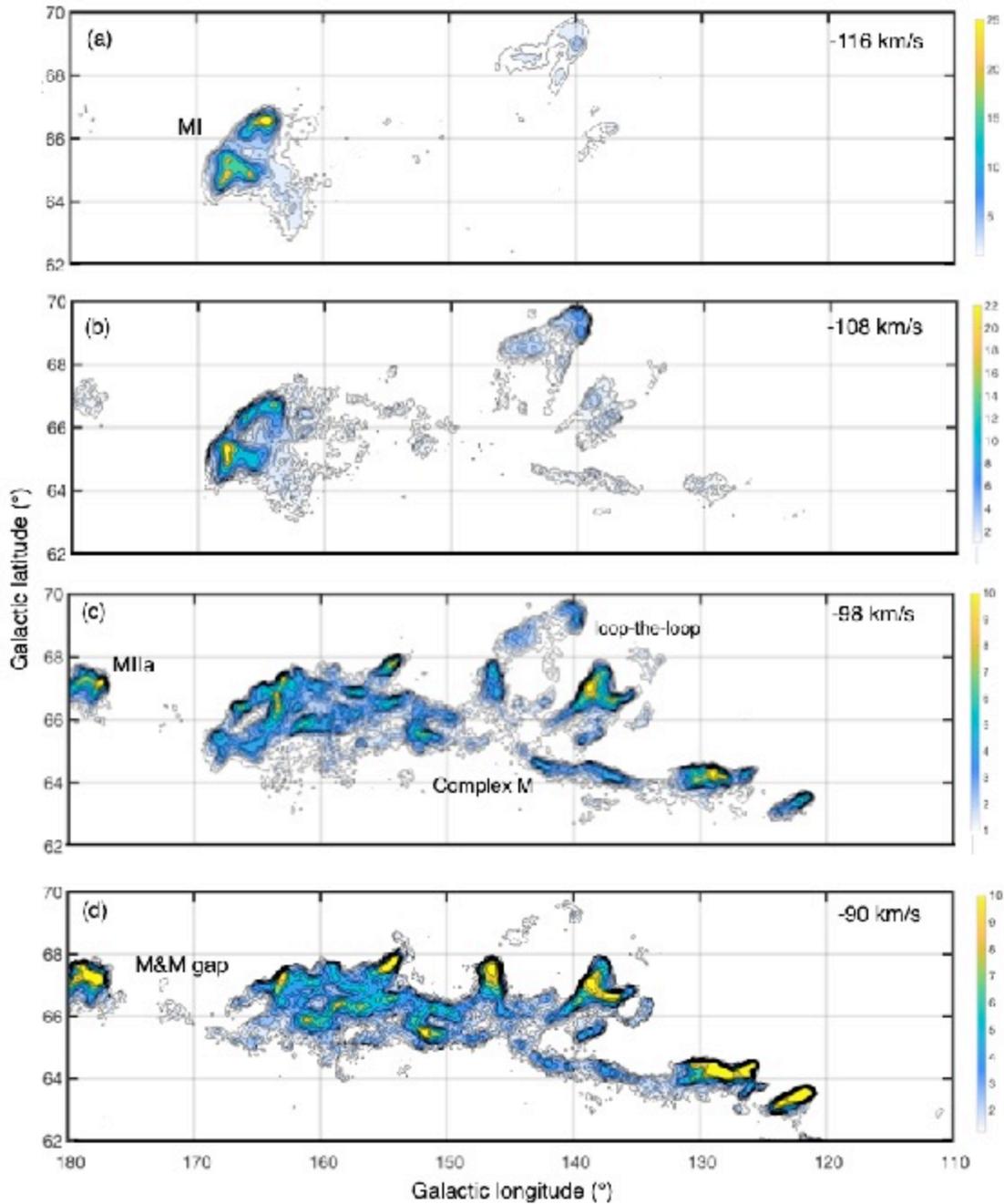

Fig 2. Close-up area maps using EBHIS data with a 2 km/s band width showing how the structure of MI changes with velocity. (a) The well-known two-peaked appearance of MI at -116 km/s; (b) MI forms a ring-shaped structure and extends to the right at -108 km/s; (c) The complete ring connects to the extended filamentary Complex M feature at -98 km/s. (d) The emission from MI has blended with Complex M; note the M&M gap between MI and MIIa. Legends are in degrees K

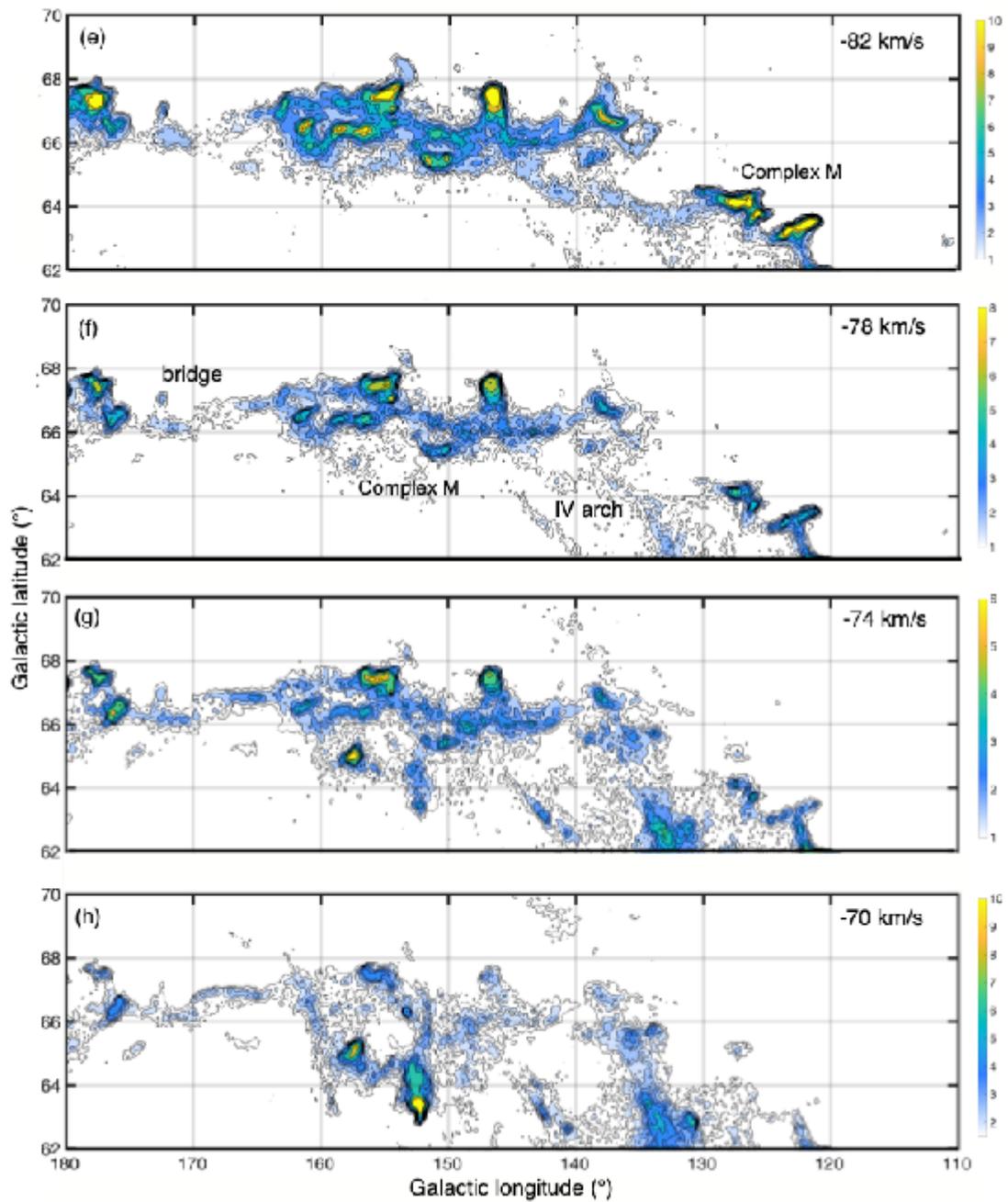

Fig 2. (e) At -82 km/s a narrow filament of gas creates a bridge between MI and MIIa; This bridge continues in frame (f) at -78 km/s and strengthens in frame (g) at -74 km/s, before it starts to fade in frame (h) at -70 km/s. The intrusive emission from the IV Arch at about l = 132º can be seen in frames (f) to (h).

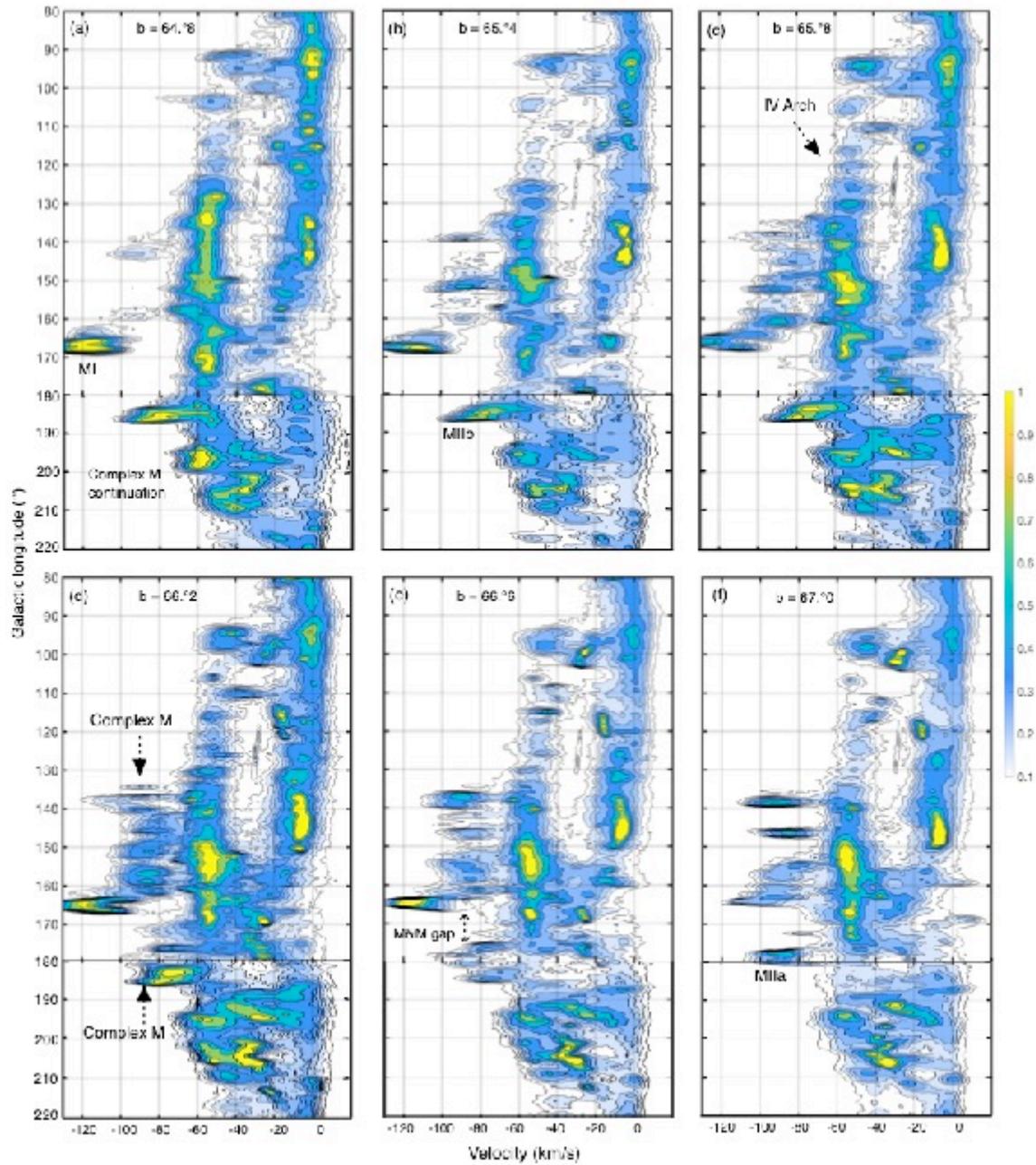

Fig 3. A series of l-v plots integrating over 0.°2 in latitude using HI4PI data that highlights the connectedness of MI and loop-the-loop segment of Complex M. (a) MI appears as an isolated cloud at b = +64°.8; (b) the connection between MI and the neighboring substructure starts to form at b = +65°.6; this connection continues to strengthen in (c) to the intensifying peaks of the loop-the-loop segment; the IV Arch is also dominant in this panel; (d) shows the strongest connection of MI to the lower-longitude gas features; (e) shows the M&M gap and the missing

low-velocity gas between l = 162 & 173°; (f) MI has blended into Complex M and here MIIa is strongest. Legends are in degrees K.

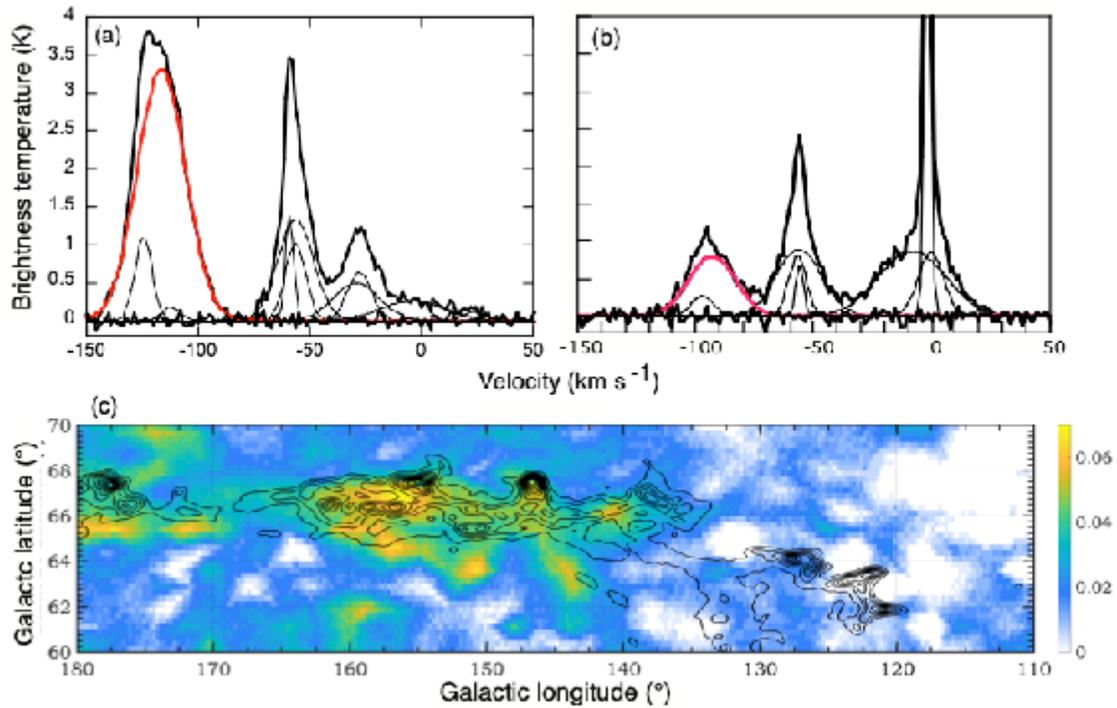

Fig 4. Comparison of HI4PI profiles and their Gauss fits toward MI and in the direction of the narrow filament segment of Complex M. (a) A profile toward MI at l = 168.°47, b = 64.°8. The line width of the broad component (red line) at -116.5 km/s is 25.4 km/s. (b) A profile toward the narrow filament in Complex M at l = 142.°5, b = 64.°5. The line width of the broad component (red line) at -93.2 km/s is 23.5 km/s. The other Gaussians required to fit the profiles are shown as thin, black lines. c) The HI4PI-data used to produce the l-b area map for Complex M showing the brighter HI contours overlaid on the H\alpha image in color (legend in Rayleighs), both at the velocity of -80 km/s with a 10 km/s-wide band. The HI contours are from 5 to 40 K in 5 K intervals.

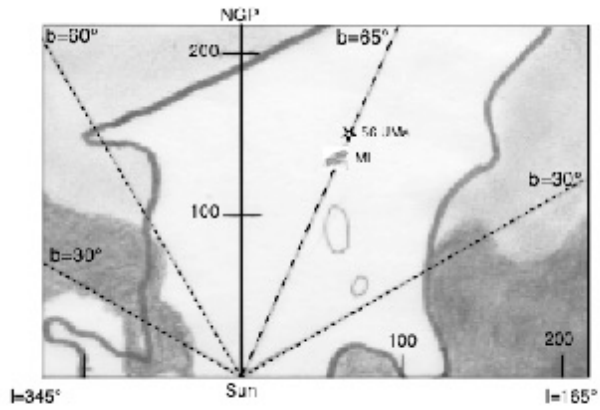

Fig 5. The outlines of the Local Chimney adapted from Fig. 8 of Lallement et al. (2003). This is the cross-section at l = 165° with distances along the two axes shown in pc. The distance to MI and the star 56 UMa are located according the scenario suggested by Schmelz & Verschuur (2022).

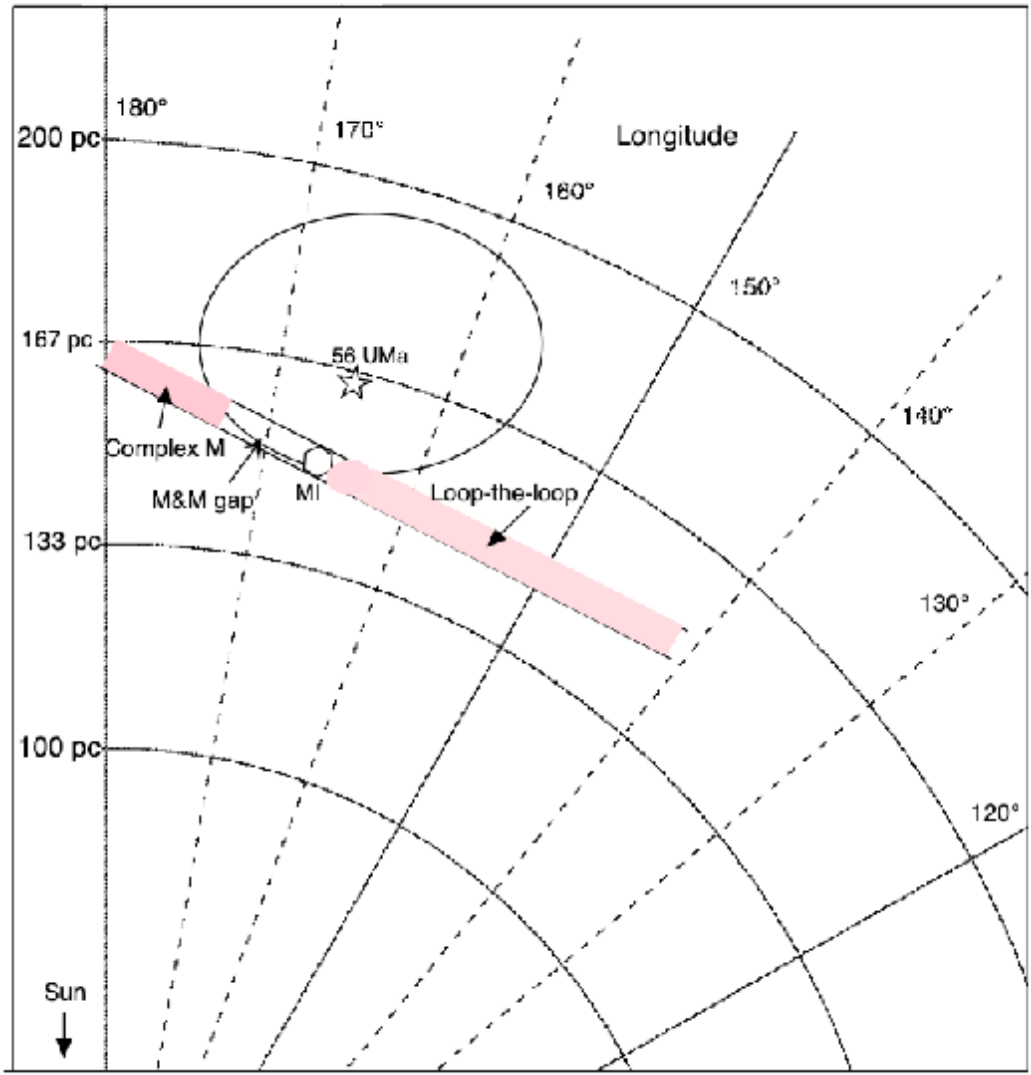

Fig 6. A polar plot schematic centered on the Sun showing a cross-section of the area at latitude of 65° with distance (in pc) versus galactic longitude. Indicated to scale are the location of yellow giant star, 56 UMa, the boundary of the supernova shell, and the current position of MI. Because the far side of the shell would be expanding onto the Local Chimney we have no a priori knowledge of its boundary in depth. The loop-the-loop segment of the Complex M arched filament at this latitude is depicted as the pink band, tilted from the plane of the sky at an angle that corresponds closely with the direction of local interstellar magnetic field direction (see text). The geometry accounts for the connection between MI and the loop-the-loop segment at lower longitudes (to the right of MI in the plot) and the M&M gap (to the left of MI). This gap is where the supernova shell has apparently interacted with and removed the neutral gas. MIIa is located at l = 178° at a latitude above this cross section where Complex M continues to higher longitudes.